\documentclass[review,12pt]{elsarticle}

\usepackage{float}
\usepackage[toc,page]{appendix}
\usepackage[scaled]{helvet}
\usepackage{nccmath}
\usepackage{lineno}   
\usepackage{epsfig}
\usepackage{amssymb}
\usepackage{amsmath}
\usepackage{upgreek}
\usepackage{booktabs}
\usepackage{lscape}
\usepackage{siunitx}
\usepackage{gensymb}
\usepackage{array}
\usepackage{makecell}
\usepackage{mathtools}
\DeclareMathAlphabet\mathbfcal{OMS}{cmsy}{b}{n}
\usepackage{hyperref}
\usepackage{algpseudocode}
\hypersetup{
	colorlinks=true,
	allcolors = blue,
}
\usepackage{longtable}
\usepackage{graphicx}
\usepackage {epstopdf}
\usepackage{psfrag}
\usepackage{mathrsfs}
\usepackage{latexsym}
\usepackage{caption}
\graphicspath{{./Figures/}}
\usepackage{xcolor}
\usepackage{subcaption}
\usepackage{setspace}       
\usepackage[normalem]{ulem} 
\allowdisplaybreaks

\begin{document}
\sloppy

\begin{frontmatter}

\title{Particulate analysis of post-thermal runaway soot of Li-ion battery using X-ray computed tomography}
\author{Avtar Singh\corref{cor}}
\ead{Avtar.Singh@nlr.gov}
\author{Donal P. Finegan}
\cortext[cor]{Corresponding author.}
\address{Center for Energy Conversion and Storage Systems, National Laboratory of the Rockies, Golden, CO 80401, United States of America}
%
\begin{abstract}
The microstructural characteristics of soot generated during thermal runaway events in lithium-ion batteries have been examined, with a focus on the associated health risks. X-ray computed tomography (Nano-CT) offers advanced 3D imaging for accurate analysis of particle shape, size, and distribution. Its primary benefit is the detection of core-shell structures that contain thin shell of carcinogenic materials. In contrast, traditional techniques like X-ray diffraction can only quantify the mass fraction of these hazardous phases, which is often minimal. Nevertheless, even a small mass fraction of carcinogenic materials on carbonaceous surfaces poses significant health risks when inhaled, potentially compromising internal organ function. This research underscores the need for a deeper understanding of particulate composition to inform safety regulations and respiratory protection measures.
\end{abstract}
\end{frontmatter}
 \section{Introduction}
Although Li-ion batteries are widely used in automotive and energy storage systems due to their high energy density and long cycle life, safety concerns remain significant~\cite{chen2021review,yao2025comprehensive}. One of the main safety issues is thermal runaway (TR), which can occur due to electrical, thermal, or mechanical abuse~\cite{song2024fault}. Internal defects within the battery can also lead to thermal runaway, even without external abuse. For example, microscopic flaws in the separator that isolates the anode and cathode can cause internal short circuits. Similarly, welding defects, electrode damage, and metal contamination can also lead to internal short circuit~\cite{qian2021role,abraham2023safe}. When an internal short circuit occurs, the battery's temperature and pressure can rise rapidly, potentially resulting in combustion or explosion, and releasing harmful gases and aerosols~\cite{chombo2020review,chen2021review,xia2023safety}.

There have been several accidents involving electric vehicles~\cite{chombo2020review}, aircraft~\cite{FAA_report_2026}, and mining operations~\cite{Mining_report_2024}, as well as fires in battery manufacturing plants~\cite{Manufac_plant_fire_1,Manufac_plant_fire_2,Manufac_plant_fire_3}, all of which have resulted in the release of harmful gases and aerosols. Notably, there have been tragic fire incidents in electric vehicles~\cite{chombo2020review}. 
From January 2006 to 2026, the U.S. Federal Aviation Administration recorded over 700 events, many of which resulted in exposure to aerosols and gases emitted by lithium-ion batteries~\cite{FAA_report_2026}. The mining industry has been adopting lithium-ion batteries in an effort to replace high-emission diesel-powered equipment used in underground operations. However, as the use of lithium-ion batteries increases in underground mines, the risk of explosions due to thermal runaway also rises~\cite{dubaniewicz2013lithium,Mining_battery_safety_guidelines_2025}.

Numerous studies have been conducted to investigate the flammability and toxicity of gases released during thermal runaway~\cite{fernandes2018identification,jia2022analysis}. Researchers have also examined the morphology and elemental composition of the soot or aerosols produced in this process using scanning electron microscopy and energy-dispersive X-ray spectroscopy~\cite{barone2021lithium}. However, to gain a deeper understanding of the size, morphology, and composition of the particulate matter ejected during thermal runaway, obtaining contrast-specific three-dimensional images would be beneficial.

In this study, we have conducted a microstructural analysis of the particulate ejecta generated during the thermal runaway of lithium-ion cells using high-resolution three-dimensional imaging with a X-ray nano-computed tomography instrument. Such research is vital for providing first responders with critical information regarding the appropriate types of personal protective equipment (PPE), particularly masks, required during firefighting operations. Furthermore, it emphasizes the PPE requirement and potential risks associated with the subsequent site cleanup after such incidents.
\section{Methods}
\begin{figure}[!htb]
	\centering
	\includegraphics[width=1.0\linewidth]{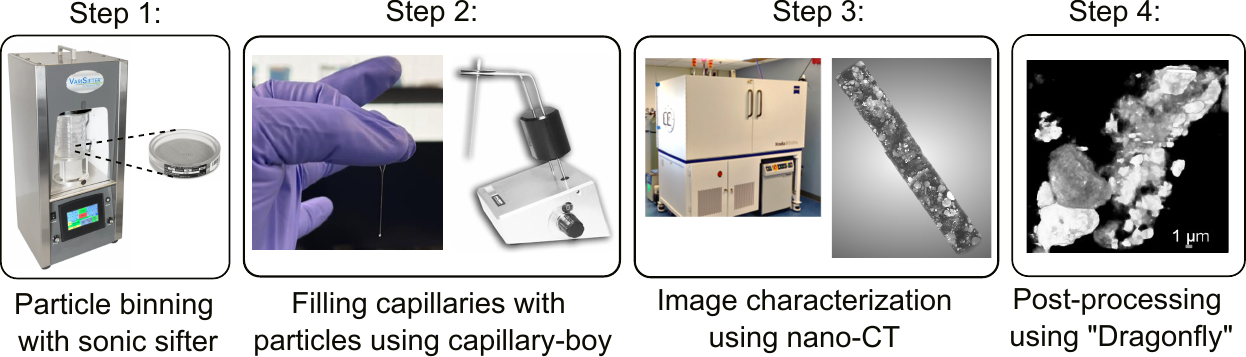}
	\caption{Workflow to characterize the microstructure of soot generated during a thermal runaway event in a lithium-ion cell, using X-ray computed tomography.}
\label{fig:TR_particle_characterization_workflow}
\end{figure}

An 18650 commercial Li-ion cell (LG M36) at 100\% state of charge was brought to thermal runaway via heating inside a fractional thermal runaway calorimetry (FTRC)~\cite{walker2019decoupling}. The LG M36 cell contained a mixed cathode of $\rm LiNi_{0.86}Co_{0.12}Al_{0.02}O_2$ (NCA) and $\rm LiMn_2O_4$ (LMO) (95:5 wt\%) and a graphite anode~\cite{iannello2019performance}. The soot was collected from the FTRC following the thermal runaway test. Fig.~\ref{fig:TR_particle_characterization_workflow} illustrates the workflow for the three-dimensional microstructural characterization of soot collected after a thermal runaway event in lithium-ion batteries. In Step 1, the soot from the thermal runaway event was collected and separated using an AdvancedTech Varisifter sonic separator to effectively isolate particles smaller than 20$\mu$m, enabling a focused analysis of the finer components of the soot. To enable detailed characterization of the images, the processed particles were carefully loaded into quartz capillary tubes with an outer diameter of 100$\mu$m, a wall thickness of 10$\mu$m, and a length of 80 mm. The particles were loaded with the assistance of a device known as a "capillary boy," as illustrated in Step 2 of Fig.~\ref{fig:TR_particle_characterization_workflow}. For the characterization of the microstructure of the thermal runaway soot contained within the capillaries, a lab-based X-ray computed tomography instrument (Zeiss Xradia 810 Ultra, Carl Zeiss) was employed. This instrument utilized a quasi-monochromatic X-ray energy of 5.4 keV to gather three-dimensional, high-resolution images of the internal structure of the soot (Step 3). Images were captured at a pixel binning of 2, resulting in a pixel size of 128 nm, which provided a field of view of 64$\mu$m × 64$\mu$m. The samples were rotated through 180 degrees, with radiographs collected at 0.2-degree intervals, yielding 901 projections. Multiple scans using the absorption contrast technique are conducted to quantify the complexity of soot resulting from thermal runaway. In the final step, the Dragonfly software was utilized for post-processing to extract detailed images and thoroughly evaluate the observed microstructural characteristics in 3D.
\section{Results and discussion}
\begin{figure}[!htb]
	\centering
	\includegraphics[width=1.0\linewidth]{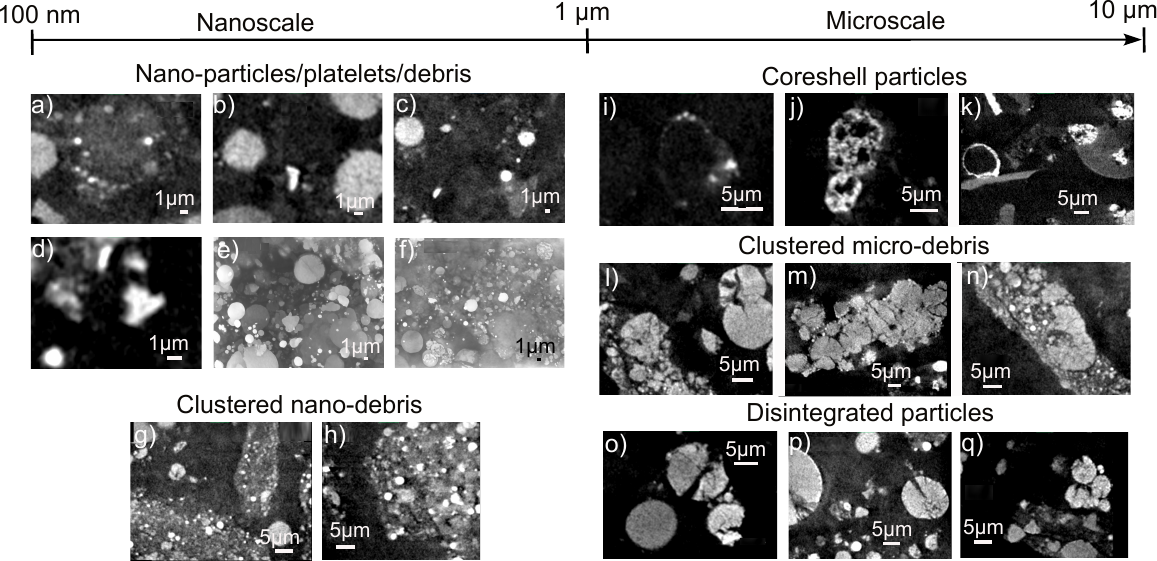}
	\caption{Classification of particulates based on the X-ray computed tomography imaging of soot generated during a thermal runaway event in a Li-ion cell. All images are 2D slices of 3D reconstructed microstructure.}
	\label{fig:TR_micro_chris_sample}
\end{figure} 
Fig.~\ref{fig:TR_micro_chris_sample} illustrates the detailed classification of particulates obtained from the nano-CT imaging technique. The analysis reveals a wide distribution of particle sizes, ranging from 100 nm to 1$\mu$m (nanoscale) and from 1$\mu$m to 10$\mu$m (microscale). Based on size alone, international conventions for respirable aerosols define penetration efficiency, which refers to the effectiveness with which particles of specific size ranges are sampled to evaluate relevant exposures~\cite{cen1993481,international2006air,dahlstrom2014acgih}. The table below lists the particle sizes and their corresponding penetration efficiencies:
\begin{itemize}
    \item At 10 $\mu$m: 1\% PE
    \item At 5 $\mu$m: 30\% PE
    \item At 3 $\mu$m: 74\% PE
    \item At 100 nm: 99.79\% PE
\end{itemize}
We observe various particle sizes, shapes, and clusters in the nano-scale range of 100 nm to 1 $\mu m$. The penetration efficiencies for these particles are significantly higher, typically exceeding 74\%, which enables them to efficiently penetrate the lungs and impact respiratory health. In the micro-scale range, various higher Z-elements (brighter greyscale due to their higher attenuation) adhere to lighter materials (darker greyscale due to their lower attenuation). While lighter materials, such as graphite at the micro-scale, may not pose life-threatening concerns, the transition metals adhering to their surfaces (in partial and complete core-shell structures in Fig.~\ref{fig:TR_micro_chris_sample}(i-k)) can create serious health hazards if inhaled or come into contact with the skin or internal surfaces within the human body. Although the volume fraction of these transition metal oxides per particle is significantly lower, they pose a high health risk. The compositions of these particles are derived from the cathode and anode materials of lithium-ion cells that underwent thermal runaway. These particles include nickel (Ni), cobalt (Co), copper (Cu), aluminum (Al), and graphite (Gr), among others. Some of the risks associated with inhaling these elements are mentioned here. The presence of transition metals greatly influences toxicity. When mixed with carbonaceous particles, transition metals can facilitate the production of reactive oxygen species, leading to oxidative damage, including DNA strand breaks and inflammation~\cite{donaldson1997free,valavanidis2000generation,wilson2002interactions,jimenez2000activation}.  For instance, exposure to as little as 0.005 $mg/m^3$ of cobalt can lead to asthma, pneumonia, and wheezing~\cite{atsdr2004atsdr}. Inhalation of nickel aerosols (>0.02 $mg/m^3$) is associated with lung inflammation and cancer~\cite{gates2023nickel,seilkop2003respiratory}. Additionally, inhaling aluminum can cause persistent coughing and negatively affect the nervous system. Carbon- and fluorine-containing aerosols also have adverse health effects~\cite{barone2021lithium}. A lack of awareness or information creates challenges in identifying hazards and determining effective respiratory protection strategies. Thus, it is crucial to systematically and carefully analyze the particulate composition and morphology of soot generated during the thermal runaway scenario of batteries.

\section{Conclusions}
The present study elucidates the complex nature of soot generated during thermal runaway events in lithium-ion batteries. X-ray computed tomography provides valuable three-dimensional imaging that enables precise analysis of particle shape, size, and distribution, as well as phase contrast between different species based on their absorption coefficients. A notable advantage of this technique is its ability to identify core-shell structures through brightness contrast arising from the varying absorption properties of materials. Despite the relatively small mass fraction of carcinogenic materials on the carbonaceous surface of the core-shell structure, their inhalation poses significant health risks, potentially affecting internal organs. The findings underscore the necessity for systematic analysis of particulate emissions to enhance awareness and inform the development of effective respiratory protection strategies. A thorough understanding of the composition and behavior of these particles is critical to establishing safety guidelines that mitigate health risks in both occupational and public settings.

\section*{Acknowledgment}
This work is authored by the National Laboratory of the Rockies, operated under Contract No. DE-AC36-08GO28308. Financing is provided by the US DOE Advanced Research Projects Agency-Energy (ARPA-E) (award number DE-AR00001723, work authorization number 22/CJ000/07/03). We thank technology manager, H. Cheeseman for their support throughout this project. The views expressed in the article do not necessarily represent the views of the DOE or the US Government. The US Government and the publisher, by accepting the article for publication, acknowledges that the US Government retains a nonexclusive, paid-up, irrevocable, worldwide license to publish or reproduce the published form of this work, or allow others to do so, for US Government purposes.

\biboptions{sort&compress}
\bibliographystyle{unsrt}
\footnotesize \bibliography{reference.bib} 
\end{document}